\documentclass[letterpaper,prl,aps,showpacs,floatfix,twocolumn]{revtex4-1}
\usepackage{graphicx}
\usepackage[ansinew]{inputenc}
\usepackage{array}
\usepackage{color}
\usepackage{amsmath}
\usepackage{amsxtra}
\usepackage{amstext}
\usepackage{amssymb}
\usepackage{latexsym}
\usepackage{dsfont}
\usepackage{verbatim}

\begin{document}

\title{Role reversal in a Bose-condensed optomechanical system}
\author{Keye Zhang$^1$, Pierre Meystre$^2$, and Weiping Zhang$^{1,}$}
\email{wpzhang@phy.ecnu.edu.cn}
\affiliation{$^1$Quantum Institute for Light and Atoms, Department of Physics, East China
Normal University, Shanghai, PRC\\
$^2$B2 Institute, Department of Physics and College of Optical Sciences, The University of Arizona, Tucson, AZ 85721, USA}

\begin{abstract}
We analyze the optomechanics-like properties of a Bose-Einstein condensate (BEC) trapped inside an optical resonator and driven by both a classical and a quantized light field. We find that this system exhibits a nature of role reversal between the matter-wave field and the quantized light field. As a result, the matter wave field now plays the role of the quantized light field, and the quantized light field behaves like a movable mirror, in contrast to the familiar situation in BEC-based cavity optomechanics [Brennecke \textit{et al}., Science \textbf{322}, 235 (2008); Murch \textit{et al}., Nat. Phys. \textbf{4}, 561 (2008)]. We demonstrate that this system can lead to the creation of a variety of nonclassical matter-wave fields, in particular cat states, and discuss several possible protocols to measure their Wigner function.
\end{abstract}

\maketitle


Optomechanics is a fast-progressing area of research that merges techniques and approaches from fields ranging from atomic, molecular and optical physics to nanoscience and to condensed matter physics. There are two major approaches to this field: a top-down approach, exploiting a range of resources from nanoscience, advanced materials, and cryogenics, and a bottom-up approach that relies largely on developments in ultracold atomic science and cavity QED. Cavity optomechanics finds its origin in ideas developed by Braginski and coworkers in the context of the interferometric detection of gravitational waves \cite{Braginsky1995}. Its more recent focus is directed in large part to the challenge of operating mechanical oscillators deep in the quantum regime, with a motivation ranging from fundamental physics tests to high-precision quantum metrology, and from the understanding of the quantum-classical interface to the realization of interfaces for quantum information networks \cite{Rugar2004, *Marshall2003, *Mancini2003,*Stannigel2011}. Recent highlights include the successful cooling of optomechanical systems to within a fraction of their quantum mechanical ground state \cite{NIST, *Caltech}.

Parallel developments in the bottom-up approach have used a cloud of ultracold atoms as a mechanical element interacting with light~\cite{Brennecke2008, *Murch2008, Zhou2009, *Jing2011}. In these realizations, the role of the movable mirror is played by a centroid or internal motion of an ultracold gas, for instance a Bose-Einstein condensate (BEC). A number of fundamental effects have already been demonstrated, including the onset of several quantum phase transitions as well as the quantum back-action of a single optical photon. Hybrid optical systems that couple nanoscale systems to atomic systems are also of much interest, as they combine relatively robust mechanical devices with the remarkable precision measurements available in  atomic, molecular, and optical physics science.

This paper considers a new bottom-up realization of a cavity optomechanical system where the roles of the optical and matter-wave fields are reversed: the role of the effective oscillating mirror is now played by an excitation of a mode of an optical cavity, while the usual role of light is now played by a trapped Bose condensate. In other words, the dynamics of the cavity mode, as an optomechanical mirror, is governed by the ``radiation pressure'' from a matter-wave field. 
We discuss a number of properties of that system, demonstrating that it can be utilized to generate cat states (which mean the superposition of coherent states in this Letter) and near-number states of the matter-wave field. We also draw an analogy between this system and a cavity-QED situation to show how it can be used to nondestructively detect the matter-wave field and reconstruct its Wigner function.

We consider a scalar BEC confined in a three-dimensional trap located inside a single-mode unidirectional ring high-$Q$ optical resonator. It is driven by a classical laser field of frequency $\Omega_p$ and wave vector $\mathbf{k}_p$, and the scattered light of frequency $\omega_{c}$ and wave vector $\mathbf{k}_c$ is collected along the axis of the resonator, see Fig.~1. Such a system has been realized and manipulated in recent experiments on BEC superradiance in a cavity~\cite{Slama2007, *Bux2011}.

Treating the incident laser field classically and the scattered field quantum mechanically, their interaction with the condensate is described by the Hamiltonian 
\begin{equation}
H=\hbar\int d\mathbf{r}\hat \Psi_e^{\dagger}(\mathbf{r}) \left [ \frac{%
\Omega_p}{2}e^{i(\mathbf{k}_p\cdot \mathbf{r}-\omega_p t)}+g_c \hat c\,e^{i%
\mathbf{k}_c \cdot \mathbf{r}} \right ] \hat \Psi_g(\mathbf{r) +\mathrm{h.c}.%
}  \label{H1}
\end{equation}
Here $\hat \Psi_e$ and $\hat \Psi_g$ are the field annihilation operators for atoms in the excited and ground electronic state, respectively, $\Omega_p $ is the Rabi frequency of the pumping laser, and $\hat c$ is the photon annihilation operator for the cavity mode, with vacuum Rabi frequency $g_c$.

\begin{figure}[ptbh]
\centering
\includegraphics[width=2.2in]{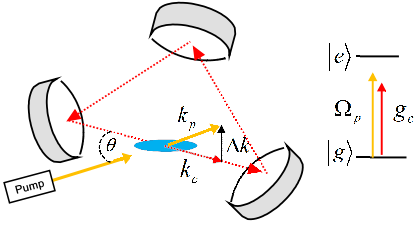}
\caption{{\protect\footnotesize {Schematic of the BEC-optical fields system and simplified level diagram of the condensed atoms, driven by a classical field of Rabi frequency $\Omega_p$ and a quantized field of vacuum Rabi frequency $g_c$.}}}
\label{scheme}
\end{figure}

For large atom-field detunings $\Delta $ it is possible to eliminate adiabatically the upper electronic state. Keeping only the lowest-order term in the scattering field, and in the rotating wave approximation, the atom-field interaction reduces then to 
\begin{eqnarray}
H &\approx &\frac{\hbar |\Omega _{p}|^{2}}{2\Delta }\int d\mathbf{r}\hat{\Psi}_{g}^{\dagger }(\mathbf{r})\hat{\Psi}_{g}(\mathbf{r})  \notag \\
&+&\frac{\hbar \Omega _{p}{g_{c}^{\ast }}}{2\Delta }\int d\mathbf{r}\left[  \hat{\Psi}_{g}^{\dagger }(\mathbf{r})\hat{\Psi}_{g}(\mathbf{r})\hat{c} ^{\dagger }e^{i\Delta \mathbf{k}\cdot \mathbf{r}}+\mathrm{h.c.}\right] ,
\label{H2}
\end{eqnarray}%
where $\Delta \mathbf{k}=\mathbf{k}_{p}-\mathbf{k}_{c}$.

\begin{figure}[ptbh]
\centering
\includegraphics[
width=2.2in ]{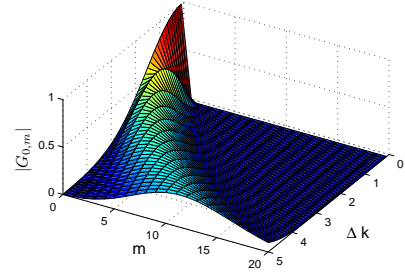}
\caption{{\protect\footnotesize {Normalized absolute value of the transition
moments $|G_{0,m}|$ between the trap ground state and its excited states as
a function of the recoil momentum $\Delta \mathbf{k}$ (in unit of $L^{-1}$). } }}
\label{Gnm}
\end{figure}

The first term in Eq.~({\ref{H2}) is a constant ac Stark shift caused by the pump field. In the following we renormalize it into the definition of the frequency of the trap potential of the ground-state atoms. The second term describes the absorption of laser light followed by scattering into the cavity mode and the reverse process, with a recoil of wave vector $\Delta  \mathbf{k}$. Expanding the field operator $\hat \Psi_g(\mathbf{r})$ in terms of the eigenstates of the three-dimensional trap, $\hat{\Psi}_g(\mathbf{r}) = \sum_{n}\hat a_n\phi_n(\mathbf{r})$, where $\hat a_n$ are the corresponding boson annihilation operators for atoms and $n\equiv (n_{x},n_{y},n_{z})$ is a triple index, that part of the interaction can be expressed as 
\begin{equation}
V = \hbar \sum_{n,m} G_{n,m}\hat a_n^\dagger\hat a_m \hat c^\dagger+\mathrm{%
h.c},
\end{equation}
where 
\begin{equation}
G_{n,m}=(\Omega_{p}g_{c}^\ast/2\Delta)\int d\boldsymbol{r}\phi_{n}^{\ast}(\boldsymbol{r} )\phi_{m}(\boldsymbol{r}) e^{i\Delta \mathbf{k}\cdot \mathbf{r }}
\end{equation}
are center of mass motional state dipole transition moments, some of which are illustrated in Fig. \ref{Gnm}. }

As discussed in Ref.~\cite{Mustecaplioglu2000}, for atoms initially in the trap ground state and large scattering angles, $\Delta \mathbf{k}\cdot \mathbf{L}\gg 1$ where $\mathbf{L}$ is the characteristic scale of the condensate in the direction of $\Delta \mathbf{k}$, one find that for short times the transition $m=0\rightarrow n>0$ dominates and triggers a superradiant scattering of the condensate. In contrast, the present work considers a situation where the condition $\Delta \mathbf{k}\cdot \mathbf{L}\ll 1$ is satisfied.  In the case, $G_{n,m}$ is sharply peaked around the transition $m=0\rightarrow n=0$, a situation similar to the Lamb-Dicke limit for trapped ions.  
For experimental setup as shown in Fig. \ref{scheme}, the condition  can easily be achieved by  a collinear arrangement for the pump and cavity field. For non-collinear case, assuming $k_p\approx k_c=2\pi/\lambda$, the condition requests the angle $\theta \ll \lambda/2\pi L$. 
As a result, a feasible angle in experiments prefers a BEC
with a size smaller than the optical wavelength.
That can be achieved by confining BEC on atom chips \cite{Purdy2010}. 

Under the condition discussed above, the single-mode approximation is safe, and we have  
\begin{equation}
V\approx \hbar G\left( \hat{c}^{\dagger }+\hat{c}\right) \hat{a}^{\dagger }\hat{a},
\label{reverse coupling}
\end{equation}
where $G\equiv G_{0,0}$ and $\hat{a}^{\dagger }\hat{a}\equiv \hat{a}_{0}^{\dagger }\hat{a}_{0}$. This is the same form as in the familiar optomechanical interaction in ultracold atomic systems~\cite{Brennecke2008,*Murch2008}, except that the roles of photons and atoms are exchanged. The optomechanical interaction is here proportional to the ``intensity'' of the matter wave $\hat{a}^{\dagger }\hat{a}$ and the ``position'' quadrature $\hat{c}^{\dagger }+\hat{c}$ of the intracavity field. This coupling can be thought of as describing the ``radiation pressure''  from a massive Schr\"odinger field driving an  ``optical oscillator'', instead of real radiation pressure from a massless optical field driving a mechanical oscillator. 
Note that here the ``position'' of the ``optical oscillator'',   as a single-mode optical field  canonical quadrature, could in principle be characterized by standard optical homodyne detection.

The effective total Hamiltonian of the light-BEC system reduces then to 
\begin{equation}
H=\hbar \omega _{0}\hat{a}^{\dagger }\hat{a}+\hbar \omega _{m}\hat{c}%
^{\dagger }\hat{c}+\hbar G\left( \hat{c}+\hat{c}^{\dagger }\right) \hat{a}%
^{\dagger }\hat{a}  \label{H4}
\end{equation}%
where the cavity-pump detuning $\omega _{m}\equiv \omega _{c}-\omega _{p}$ plays the role of the natural frequency of the ``optical oscillator''. It  can cover a wide range of values, from $\sim2\pi\times 10^2$MHz to $\sim2\pi\times 10$Hz, and can even be negative, which is impossible for a mechanical oscillator. The frequency $\omega _{0}$ is the renormalized ground-state center-of-mass frequency of the trapped BEC, typically of the order of $ 2\pi \times 10$kHz for relatively tight traps.  For $^{87}$Rb atoms, a classical Rabi frequency $\Omega _{p}=2\pi \times 1$GHz, $g_{c}=2\pi \times 10$MHz, and $\Delta =2\pi \times 5$GHz, we find $G \approx 2\pi \times 1$MHz.  As in the usual optomechanical interaction, the sign of the optomechanical coupling $G$ can be changed by changing the sign of cavity detuning, in this case the atom-field detuning $\Delta $. 

In the usual optomechanics situation the movable mirror or trapped ultracold atomic system acts as a Kerr-type medium for the cavity field, resulting in a series of nonlinear optical effects that include optical bistability \cite{Dorsel1983}, chaos \cite{Carmon2007}, and squeezing \cite{Fabre1994}. More generally, Ref.~\cite{Mancini1997,*Bose1997} showed that a broad variety of nonclassical states of the cavity field can be prepared by means of the optomechanical coupling. In the present situation with its reversal of roles between light and the mechanical oscillator, one finds similarly that a rich variety of nonclassical states can be generated in the BEC. In particular, in case the cavity field is initially in a vacuum state $|0\rangle_{c}$ and the BEC in a coherent state $|\alpha\rangle_{a}$ we find that in the interaction picture the final state of the system is 
\begin{equation}
|\Psi(t)\rangle =e^{-\frac{|\alpha|^2}{2}}\sum_{n=0}^\infty \frac{\alpha^n}{\sqrt{n!}}e^{i\Lambda^2n^2(\omega_mt-\sin \omega_m t)}| n\rangle_a \otimes|\Lambda n\eta\rangle_c,  \label{S1}
\end{equation}
where $|n\rangle _{a}$ are number states of the BEC, $|\Lambda n \eta \rangle _c$ are coherent states of the cavity field of complex amplitude $\Lambda n\eta$, with $\eta = 1 - \exp(-i\omega_m t)$, and we have introduced the coefficient $\Lambda \equiv -G/\omega_m$. In the resolved sideband limit of cavity optomechanics, $\omega_m \gg \kappa$ where $\kappa$ is the resonator width, $\Lambda > 1$ would correspond to the strong-coupling regime, where a single photon substantially modifies the resonator properties. In the present case this parameter is likewise a measure of the nonlinearity of the dynamics, as we shall see. But in contrast to the usual situation, the ``strong coupling regime'' can now be easily reached by simply increasing the Rabi frequency $\Omega_p$ of the classical driving field.

At times $\omega _{m}t=2m\pi $, $m$ integer, the state of the cavity field returns to the vacuum while the state of BEC can take the form of various cat states, depending on the value of $\Lambda $. For instance, when $\Lambda ^{2}=1/4m$, the BEC is in the two-component cat state 
\begin{equation}
|\phi \rangle _{a}=\frac{1+i}{2}|\alpha \rangle _{a}+\frac{1-i}{2}|-\alpha \rangle _{a}.  \label{cat}
\end{equation}
The choice of other values of $\Lambda $ can also lead to the generation of multicomponent cat states \cite{Bose1997}.

Except at those times, the state (\ref{S1}) is an entangled state of the cavity field and the BEC. The entanglement is most pronounced at $\omega _{m}t=(2m+1)\pi $, providing a route to the preparation of nonclassical state of the condensate via conditional measurements on the cavity field. In particular, a measurement of the quadrature $\hat{X}=(\hat{c}+\hat{c}^{\dagger })/2$, that is, the ``position'' of the equivalent mirror will force the BEC into a state 
\begin{equation}
|\tilde{\phi}\rangle _{a}\propto e^{-\frac{\left\vert \alpha \right\vert ^{2}}{2}}\sum_{n=0}^{\infty }\frac{\alpha ^{n}}{\sqrt{n!}}e^{i(2m+1)\Lambda^{2}n^{2}\pi }f_{n}(X)|n\rangle _{a},
\end{equation}
where the number state distribution is dominates by the scalar products 
\begin{equation}
|f_{n}(X)|^{2}=|\langle X|2\Lambda n\rangle _{c}|^{2} \propto \exp [-8\Lambda^{2}(n-X/2\Lambda )^{2}].
\end{equation}
Provided that $X/2\Lambda$ is near an integer $m$ and Gaussian width $\sigma =1/4\Lambda \ll 1$, 
the coefficients $|f_n(X)|^{2} (n\neq m)$ are then strongly suppressed so that the state $|\tilde{\phi}\rangle _{a}$ is well approximated by  the  number state $|m \rangle _{a}$.

The preparation of near number states via conditional measurements of the cavity field provides a indirect way to extract the atom number statistics of the condensate. However the number statistics alone are not sufficient to verify the emergence of atomic Schr{\" o}dinger cat states. To gain the necessary phase information, one can adapt an optical method that involves the mixing of the field to be characterized with a reference classical field with an adjustable relative phase~\cite{Yurke1990}. This can be achieved in principle in a way recently demonstrated~\cite{Gross2011} for the atomic homodyne detection of entangled twin-atom states in a spinor BEC. The idea is to mix the state to be characterized with another macroscopically populated atomic state that serves as a reference, for instance via a microwave field induced transition. The relative phase between the two atomic states can be adjusted through the phase of the microwave field.

\begin{figure}[ptbh]
\centering
\includegraphics[
width=3.1in
]{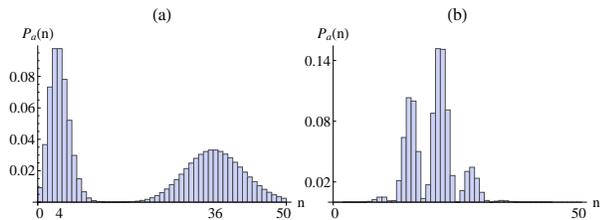}
\caption{{\protect\footnotesize {\ Atom number distribution $P_{a}(n)$ for
the displaced cat state (\protect\ref{cat}), with $|\protect%
\alpha|=2$ and $|\protect\beta|=4$. (a) $\protect\alpha$ and $\protect\beta$
are in phase. (b) $\protect\alpha$ and $\protect\beta$ are $\protect\pi/2$
out of phase. }}}
\label{bar}
\end{figure}

In one possible measurement protocol, the optomechanical interaction $G$ is rapidly switched off at a time $t$ when the cavity field is back in the vacuum and the condensate in a Schr{\''o}dinger-cat state, simply by switching off the classical driving field of Rabi frequency $\Omega_p$ in a time short compared to the BEC decoherence time. The BEC atoms are then coupled by a microwave field to another, macroscopically populated atomic state. As a result the density operator of the system becomes $\hat{\rho}=\hat{D}(\beta)  \hat{\rho}_a\hat{D}(-\beta) \otimes|0\rangle\langle0|_c$, where the atomic displacement operator $\hat{D}(\beta)$ describes the mixing with a reference field of complex amplitude $\beta$. Turning then the optomechanical coupling back on with $\Lambda \gg 1/4$, we find that after a time $\omega_m t=\pi$, the probability of getting the value $X$ for the ``position'' quadrature of the optical field is given by 
\begin{equation}
P_c(X) =\langle X| \mathrm{Tr}_a[\hat {U}(\pi) \hat{\rho}\hat{U}^{-1}(\pi)
]|X\rangle=\sum_nP_a(n) |f_n(X)|^2,
\end{equation}
where Tr$_{a}$ is a partial trace on the atomic system, $P_a(n)={_a\langle} n|\hat{D}(\beta) \hat{\rho}_a\hat{D}(-\beta)| n\rangle _a$ is the atom number statistics for the displaced matter-wave field, and $\hat{U}(t)$ is the time evolution operator.
Figs.~\ref{bar}(a) and \ref{bar}(b) show the in-phase and out-of-phase values of $P_a(n)$ for $|\alpha| =2$ and $|\beta|=4$. In the in-phase case, $P_a(n)$ is the sum of two quasi-Poissonian distributions peaked around $|\alpha+\beta|^2$ and $|-\alpha + \beta|^2$. When $\alpha$ and $\beta$ are $ \pi/2$ out of phase, the interference between the two atomic fields results in $P_a(n)$ exhibiting a Poisson envelope with superimposed modulations.

A variation on a protocol inspired by cavity QED experiments~\cite{Del'eglise2008} allows to reconstruct the full Wigner function of the field. To see how this works, we start from the general expression of the Wigner function of a single-mode field in the number state basis~\cite{Cahill1969}, 
\begin{equation}
W_a(\beta)=\frac{2}{\pi}\sum_{n=0}^{\infty}(-1)^{n}{_{a}\langle}n|\hat{D}(-\beta) \hat{\rho}_a\hat{D}(\beta)|n\rangle_a,  \label{W}
\end{equation}
where the $(-1)^n$ factor comes from the expectation value of the number parity operator.

To determine the Wigner function~(\ref{W}) through a measurement of the cavity field, we again assume that the system is in the uncorrelated initial state $\hat{\rho}_{a}\otimes{\left\vert 0\right\rangle \left\langle 0\right\vert }_{c}$, and perform first a displacement of the matter-wave field before switching on the classical driving field of Rabi frequency $\Omega_p$. The probability of finding the cavity field in the $n$-photon state is then found to be 
\begin{equation}
P_c(n) =\sum_{m=0}^{\infty} \left ( \frac{| \Lambda \eta m|^{2n}}{n!} \right
) e^{-|\Lambda \eta m|^2} {_{a}\langle m|}\hat{D}(\beta) \hat{\rho}_{a}\hat{D%
}(\beta){|m\rangle}_{a}.
\end{equation}
Comparing this result with the expression (\ref{W}) shows that by chosing $|\Lambda \eta|^2=\pi$ we have simply 
\begin{equation}
W_a(\beta)=\frac{2}{\pi}\sum_{n=0}^{\infty}P_{c}\left( n\right) (1+i)^n.
\end{equation}
That approach is similar in spirit to a method used in Ref. \cite{Del'eglise2008} to measure the Wigner function of the intracavity microwave field in a micromaser via detection of the state of the atoms that drive that field. One disadvantage is that it relies on the acquisition of a large amount of experimental data as well as on the the need to resolve the exact photon number $n$.

An alternative method that yields a direct measurement of $W_{a}(\beta)$ can be implemented if the cavity contains a single photon, a situation that might be achieved using a photon blockade effect, maybe by including in addition to the off-resonant condensate atoms a single resonant atom coupled to the light field in the strong-coupling regime of cavity QED \cite{Birnbaum2005}, or perhaps via a movable mirror \cite{Rabl2011}. While this would increase significantly the complexity of the experiment, the advantage of that approach is that the Wigner function is a directly measured quantity~\cite{Lutterbach1997}.

With the cavity field initially in a mixture $\hat{\rho}_{c}=\rho_0|0\rangle \langle0| + \rho_1|1\rangle \langle 1|$, displacing the matter wave field as before by an adjustable amount $\beta$, and then switching on the optomechanical interaction, we find that the difference in probabilities $\Delta P_c = P_c(1)-P_c(0)$ of measuring no photon and one photon in the resonator is 
\begin{equation}
\Delta P_c= (\rho_1-\rho_0)\sum_{n=0}^\infty {_a\langle} n|\hat D(-\beta)\hat \rho_a \hat D(\beta) \cos(2 \Lambda|\eta|n)|n\rangle_a.
\end{equation}
By adjusting the parameter $\Lambda$ and/or the interaction time so that $2\Lambda |\eta| =\pi$, the atomic Wigner function takes the simple form 
\begin{equation}
W_a(\beta)=\frac{2 \Delta P_c}{\pi(\rho_{1}-\rho_{0})}.
\end{equation}

In summary we have investigated a new configuration of a BEC-based cavity optomechanical system characterized by a role reversal of the light and matter-wave fields as compared to the usual situation. We showed how this system can not only efficiently prepare nonclassical states of the BEC, but can also nondestructively characterize them and even reconstruct their Wigner function. This extends the research of cavity optomechanics into the realm of quantum matter wave optics and opens the way to a rich new direction of investigations, with potential in quantum information processing and quantum metrology. The proposed schemes are experimentally challenging, and a number of issues need to be addressed, from the quantum efficiency of the detectors to technical and fundamental noise issues.  

One advantage of the optomechanical system considered here over conventional one is that the ``optical oscillator'', a single-mode optical field in a high-Q cavity, is immune to the thermal and clamping noises. It is only subject to the vacuum  fluctuations, an effective zero temperature reservoir. The condensate can also be considered to be effectively at zero temperature, so the main source of environment decoherence are expected to be three-body collisions, which result in fluctuations in the atom number $\langle \hat a^\dagger \hat a \rangle$, shot noise due to cavity losses,  and the intensity and phase fluctuations of the classical driving field $\Omega_p$. (We remark that for quantum-degenerate atomic systems $s$-wave collisions are a coherent nonlinear mechanism that does not result in decoherence.)  Since the quantum states of photons and atoms are highly entangled, these decoherence mechanisms are expected to have nontrivial effects on the generation of nonclassical atomic states \cite{Mancini1997,*Bose1997}.  The role of these and other decoherence and noise mechanisms will be the object of a detailed future paper.

This work is supported in part by the National Basic Research Program of China under Grant No. 2011CB921604 (W. Z), the National Natural Science
Foundation of China under Grant No. 11004057, the Fundamental Research Funds for the Central Universities (K. Z), the the US National Science Foundation, the DARPA QuASAR and ORCHID programs through grants from AFOSR and ARO, and the US Army Research Office (P. M).

\bibliographystyle{apsrev4-1}
\bibliography{reoptomech}

\end{document}